\documentclass[10pt]{article} 

\usepackage{graphicx, pdfpages}
\usepackage{latexsym}
\usepackage{mathtools}
\usepackage{amsmath, amsthm, amssymb}

\newcommand{\be}{\begin{equation}}
\newcommand{\ee}{\end{equation}} 
\newcommand{\lb}{\label}
\newcommand{\OL}{\overline}

\newcommand{\ba}{{\bf a}}

\newcommand{\bk}{{\bf k}}

\newcommand{\br}{{\bf r}}
\newcommand{\bu}{{\bf u}}

\newcommand{\bx}{{\bf x}}

\newcommand{\bJ}{{\bf J}}

\newcommand{\wt}{\widetilde}

\newcommand{\grad}{{\mbox{\boldmath $\nabla$}}}
\newcommand{\bdot}{{\mbox{\boldmath $\cdot$}}}

\newcommand{\bzed}{{\mbox{\boldmath $0$}}}

\begin{document}

\baselineskip=18pt
\begin{center}
\begin{LARGE}
{\bf The conservative cascade of kinetic energy in compressible turbulence}\\
\end{LARGE}

\bigskip
\bigskip

Hussein Aluie$^{1,2}$, Shengtai Li$^{2}$, and Hui Li$^{3}$\\
{\it
  $^{1}$ Center for Nonlinear Studies\\
$^{2}$ Applied Mathematics and Plasma Physics (T-5)\\
$^{3}$ Nuclear and Particle Physics, Astrophysics and Cosmology (T-2)\\
Theoretical Division, Los Alamos National Laboratory, Los Alamos, New Mexico 87545, USA}

\bigskip
\bigskip

\begin{abstract}
The physical nature of compressible turbulence is of fundamental importance in
a variety of astrophysical settings.
We present the first direct evidence that mean kinetic energy cascades conservatively
beyond a transitional ``conversion'' scale-range despite not being an invariant of the 
compressible flow dynamics. 
We use high-resolution three-dimensional simulations of compressible hydrodynamic turbulence 
on $512^3$ and $1024^3$ grids. We probe regimes of forced steady-state isothermal flows and 
of unforced decaying ideal gas flows. The key quantity we measure is pressure dilatation cospectrum,
$E^{PD}(k)$, where we provide the first numerical evidence that it decays at a rate faster than $k^{-1}$ as a function of wavenumber. 
This is sufficient to imply that mean pressure dilatation acts primarily at large-scales and that
kinetic and internal energy budgets statistically decouple beyond a transitional scale-range. 
Our results suggest that an extension of Kolmogorov's inertial-range theory to compressible turbulence
is possible.
\end{abstract}
\end{center}

\vspace{0.5cm}

~~~{\bf Key Words:} hydrodynamics --- methods: numerical --- turbulence.
\clearpage

\section{Introduction}

Turbulence plays a critical role in essentially all astrophysical systems
that involve gas dynamics. For example, interstellar turbulence is believed to be driven 
on large scales by differential galactic rotation or supernovae explosions
and dissipated at the smallest scales by microphysical processes. It is widely
believed that energy is transferred from the largest scales down to the dissipation
scales through a cascade process. Measurements of interstellar scintillation (ISS)
(\cite{Armstrongetal81}) caused by scattering in the interstellar medium shows a 
power-law electron density spectrum 
with a slope close to Kolmogorov's $-5/3$ scaling over 5 decades in scale.
A composite spectrum combining ISS data with that of differential Faraday rotation angle and 
gradients in the average electron density also yields a Kolmogorov power-law over
at least ten decades in scale, dubbed as ``the great power-law in the sky'' (\cite{Armstrongetal95}).
High-resolution simulations of the interstellar medium (\cite{Kritsuketal07})
also reveal power-law scaling of spectra and structure functions which are 
reminiscent of incompressible turbulence (albeit with different slopes).
Yet, observations of the interstellar medium
suggest that the Mach number of turbulent motions is of order $0.1$ to $10$,
such that the flow is often compressible (\cite{ElmegreenScalo04}).
Similarly, high-resolution simulations of the intracluster medium (ICM) between galaxies
(\cite{Xuetal09,Xuetal10}) reveal Kolmogorov-like spectra of kinetic energy at subsonic
and transonic Mach numbers. It is not clear how Kolmogorov's 1941 phenomenology,
which forms the cornerstone for our understanding of incompressible hydrodynamic turbulence,
can carry over to such flows which exhibit significant compressibility effects.
Several recent studies have addressed the problem of compressible hydrodynamic turbulence
numerically (e.g. \cite{Porteretal98,Kritsuketal07,Schmidtetal08,Federrathetal10,PetersenLivescu10})
and theoretically (e.g. \cite{RubinsteinErlebacher97,Boldyrev02,Falkovichetal10}), but a
physical description equivalent to that of Kolmogorov's still eludes us.

The idea of a cascade itself is without physical basis since kinetic energy is not
a global invariant of the inviscid dynamics. Compressible flows 
allow for an exchange between kinetic and internal energy through two mechanisms:
viscous dissipation and pressure dilatation. While the former process is localized 
to the smallest scales just like in incompressible turbulence, 
the latter is a hallmark of compressibility and can {\it a priori} allow for an exchange 
at any scale through compression and rarefaction. 
Recently, \cite{Aluie11c,Aluie11b} showed how the assumption that 
pressure dilatation cospectrum decays fast enough rigorously implies that
mean kinetic energy cascades conservatively despite not being an invariant. 
The assumption entails that the mechanism of pressure dilatation acts primarily
at the largest scales and vanishes \emph{on average} at smaller scales.
In this Letter, we shall provide the first empirical evidence in support of this 
assumption.

The outline of this Letter is as follows. 
Section \ref{Numerics} describes our numerical simulations and 
Section \ref{Coarse-graining} presents our coarse-graining method for analyzing nonlinear scale interactions.
Section \ref{sec:PressDil} discusses the significance of pressure dilatation cospectrum and 
the assumption we are testing. Our main results are then described in Section \ref{sec:Results}
followed by a discussion on the role of shocks in Section \ref{sec:Shocks}. The Letter concludes
with Section \ref{sec:Conclusion}.

\section{Numerical simulations}\label{Numerics}

We analyze data from four numerical simulations of compressible
turbulence, summarized in Table \ref{table:Runs},
by solving the continuity, momentum, and internal energy equations
\begin{eqnarray} 
&\partial_t \rho& + \grad\bdot(\rho \bu) = 0~~, \lb{continuity} \\
&\partial_t (\rho \bu)& + \grad\bdot(\rho \bu\bu) 
= -\grad P   + \rho {\bf F}~~,   \lb{momentum}\\
&\partial_t (\rho e)&+ \grad\bdot(\rho e \bu)
= -P\grad\bdot \bu    ~~,\lb{internal-energy}
\end{eqnarray}
supplemented with an equation of state (EOS) for the fluid.
Here, $\bu$ is velocity, $\rho$ is density,  $P$ is pressure, 
$e = P/[(\gamma-1)\rho]$ is internal energy per unit mass for 
a heat capacity ratio $\gamma$, and
${\bf F}$ is an external acceleration field stirring the fluid.
We carry out two kinds of simulations: runs I and III in which the fluid is isothermal 
and the flow is constantly driven by a non-zero ${\bf F}$, and runs II and IV in which
the fluid is an ideal gas and the turbulence is decaying with ${\bf F}=\bzed$ 
(see Table \ref{table:Runs}).

The simulation domain is a periodic box ${\mathbb T}=[0,2\pi)^3$.
For isothermal forced runs I and III, we start with a uniform density field
$\rho(\bx,t=0) = 1$. The forcing function is adapted from \cite{ChoRyu09}
and has the form ${\bf F}(\bx,t) = \sum_{j=1}^{22} \hat{{\bf F}}(\bk_j) \exp{(i \bk_j\bdot\bx)} + \mbox{complex conjugate}$.
The Fourier amplitude, $\hat{{\bf F}}(\bk_j)$, has a compressive component parallel to $\bk_j$
and another solenoidal component perpendicular to $\bk_j$. 
On average, the two components are of equal magnitude. 
The amplitudes and phases of $\hat{{\bf F}}(\bk_j)$ are random in time.
The 22 forced wavevectors, $k_j$, are nearly isotropically distributed in the range 
$2\le |k_j| \le \sqrt{12}$. The initial conditions for unforced decaying Runs II and IV are 
taken from the last recorded output of Runs I and III, respectively, 
after reaching a statistically steady state.

We use the central finite-volume scheme on overlapping cells (\cite{Liuetal07})
to solve eqs. (\ref{continuity})-(\ref{momentum}) in conservative form. 
Instead of  solving internal energy eq. (\ref{internal-energy}), the code 
solves the equivalent total energy equation.
The code can also solve compressible  ideal magnetohydrodynamics.
Details of the whole algorithm have been
documented in \cite{Li08} and \cite{Li10}. Our method has been verified to
achieve the expected order of accuracy and to have very low numerical
dissipation. The high-order, low-dissipation, and divergence-free
properties of this method make it an excellent tool for simulating compressible 
hydrodynamic and magnetohydrodynamic turbulence.

Figure \ref{Fig:isoeosSpectra} plots spectra of the solenoidal and compressive
components of the velocity field from runs III and IV, both of which exhibit putative
Kolmogorov-like power-law scaling over wavenumber range $k\in [2,50]$. 
Here, the velocity field is decomposed as
$\bu(\bx) = \bu^s(\bx) + \bu^c(\bx)$, such that $\bu^s$ is solenoidal and $\bu^c$
is irrotational. Spectra from run III are averaged over times after kinetic energy 
has reached steady state. Spectra from the decaying case, run IV, are normalized 
at each time by the value of $\langle |\bu|^2\rangle$ at that time. Here, $\langle\dots\rangle$
is a volume average over the domain ${\mathbb T}$, $\int_{\mathbb T} d^3\bx (\dots)$.

\begin{table}[h]
\centering
\begin{tabular}{|ccccc|}
\hline
Run & $N^3$ & Flow & EOS & $M_{t}$\\ 
\hline
I & $512^3$ & forced & isothermal  &  0.43 \\
II & $512^3$ & decaying & ideal gas  &   -  \\
III & $1024^3$ & forced & isothermal     &  0.44    \\
IV & $1024^3$ & decaying & ideal gas  & - \\
\hline
\end{tabular}
\caption{Simulation parameters: $N^3$ is the grid-size, $M_t=u_{rms}/c_{th}$ is the Mach number
and $c_{th}$ is the sound speed. 
In the decaying runs, initial $M_t$ is similar to that of forced runs but monotonically decays in time.}
\lb{table:Runs}\end{table}
\vspace{1mm}

\begin{figure}
\centering
\includegraphics[angle=0,totalheight=.4\textheight,width=.45\textwidth]{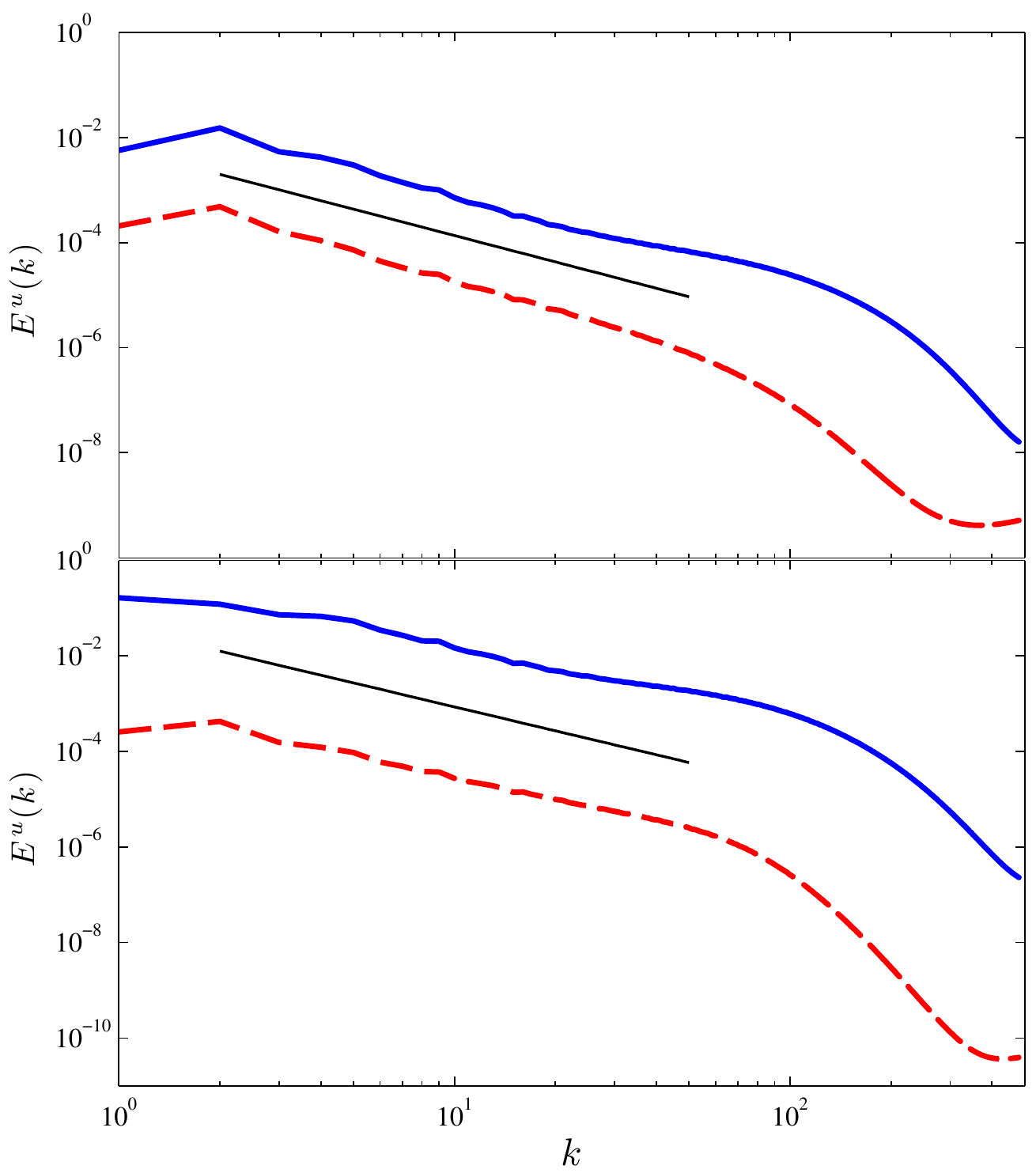}
\caption{Plots of velocity spectra from run III (top) and run IV (bottom). 
Solid (blue) plot shows spectrum of the solenoidal velocity component. 
Dashed (red) shows compressive velocity component. 
Straight lines are for reference and have a $-5/3$-slope.}
\lb{Fig:isoeosSpectra}\end{figure}

\section{Analyzing nonlinear scale interactions}\label{Coarse-graining}
The key analysis method we use is a ``coarse-graining'' or ``filtering'' approach 
common in Large Eddy Simulation (LES) literature on turbulence modelling. 
See for example \cite{Leonard74,Germano92,MeneveauKatz00} and references therein. 
The approach was developed by \cite{Eyink95a,Eyink05} to analyze 
nonlinear scale interactions in flow fields. It was further refined and utilized 
by \cite{AluieEyink09} and extended to magnetohydrodynamic (\cite{AluieEyink10}),
geophysical (\cite{AluieKurien11}), and compressible (\cite{Aluie11c,Aluie11a,Aluie11b}) flows. 
The method itself is simple. For any field $\ba(\bx)$,
a ``coarse-grained'' or (low-pass) filtered field, which contains modes
at length-scales $>\ell$, is defined as
\be
\OL \ba_\ell(\bx) = \int d^3\br~ G_\ell(\br) \ba(\bx+\br),
\lb{filtering}\ee
where $G(\br)$ is a normalized convolution kernel, $\int d^3\br ~G(\br)=1$. 
An example of such a kernel is the Gaussian function, $G(r) = \frac{1}{\sqrt{2\pi}}e^{-r^2/2}$.
Its dilation $G_\ell(\br)\equiv \ell^{-3} G(\br/\ell)$ 
has its main support in a ball of radius $\ell$. Operation (\ref{filtering}) may 
be interpreted as a local space average.

From the dynamical equation of field $\ba(\bx)$,
coarse-grained equations can then be written to describe the evolution of 
$\OL{\ba}_\ell(\bx)$ at every point $\bx$ in space and at any instant of time. Furthermore,
the coarse-grained equations describe flow at scales $>\ell$, for arbitrary
$\ell$. The approach, therefore, allows for the simultaneous 
resolution of dynamics \emph{both in scale and in space}, similar to wavelet analysis, and admits
intuitive physical interpretation of various terms in the coarse-grained balance.

Moreover, coarse-grained equations describe the large-scales whose
dynamics is coupled to the small-scales through so-called \emph{subscale} 
or \emph{subgrid} terms. These terms depend inherently on the unresolved 
dynamics which has been filtered out. 
The approach thus quantifies the coupling between different scales and may be used to extract
certain scale-invariant features in the dynamics.

\subsection{Analyzing Compressible Flows}
\cite{Aluie11c,Aluie11a} showed 
how a Favre (or density-weighted) decomposition can be employed to extend 
the coarse-graining approach to compressible turbulence. A Favre 
filtered field is weighted by density as
\be \wt{f}_\ell(\bx) \equiv \frac{\OL{\rho f}_\ell(\bx)}{\OL\rho_\ell(\bx)}. \lb{FavreDef}\ee
In the rest of this Letter, we shall take liberty of dropping subscript  
$\ell$ whenever there is no risk of ambiguity. 
The resultant large-scale dynamics for continuity and momentum are, respectively,
\be
\partial_t \OL{\rho} + \partial_i(\OL\rho \wt{u}_i) = 0.
\lb{Favrecontinuity}\ee
\begin{eqnarray} 
\partial_t \OL\rho \wt{u}_i + \partial_j (\OL\rho \wt{u}_i~\wt{u}_j )
& = & -\partial_j\left(\OL\rho~\wt\tau(u_i,u_j)\right) -\partial_i\OL{P} + \OL{\rho} \wt{F}_i, \nonumber\\ 
\lb{Favremomentum}\end{eqnarray}
where 
\be\OL\rho\wt\tau(u_i,u_j)\equiv \OL\rho(\wt{u_iu_j} - \wt{u}_i~ \wt{u}_i)\lb{Favrestress}\ee
is the \emph{subgrid stress} from the eliminated scales $<\ell$.
It is also straightforward to derive a kinetic energy budget for the large-scales, which reads
\be
\partial_t \OL\rho\frac{|\wt\bu|^2}{2} + \grad\bdot\bJ_\ell 
= -\Pi_\ell -\Lambda_\ell + \OL{P}_\ell\grad\bdot\OL\bu_\ell
+\epsilon^{inj},
\lb{largeKE}\ee
where $\bJ_\ell(\bx)$ is spatial transport flux of large-scale kinetic energy, 
and $\epsilon^{inj}(\bx)$ is the energy injected due to external stirring. 
The definition of these terms can be found in \cite{Aluie11c,Aluie11b}. 
$-\OL{P}\grad\bdot\OL\bu$ is large-scale pressure dilatation,  
and $\Pi_\ell(\bx)+\Lambda_\ell(\bx)$ is the subgrid scale (SGS) kinetic energy flux to scales $<\ell$,
\begin{eqnarray}
&\Pi_\ell(\bx)& = ~  -\OL{\rho}~ \partial_j\wt{u}_i  ~\wt\tau(u_i,u_j) ~~ \lb{flux1}\\[0.3cm]
& \Lambda_\ell(\bx)& = ~  \frac{1}{\OL\rho}\partial_j\OL{P}~\OL\tau(\rho,u_j) ~~ \lb{flux2}
\end{eqnarray}
where 
\be  \OL\tau(\rho,\bu) \equiv\OL{(\rho\bu)}_\ell-\OL{\rho}_\ell\OL{\bu}_\ell \lb{tau-def} \ee
in expression (\ref{flux2}) is the \emph{subgrid mass-flux}.
Equations (\ref{Favrecontinuity})-(\ref{largeKE}) describe the dynamics at scales $>\ell$, for arbitrary $\ell$,
at every point $\bx$ and at every instant in time. They hold for each realization of the flow without any statistical 
averaging.

The SGS flux is comprised of deformation work, $\Pi_\ell$, and baropycnal work, $\Lambda_\ell$,
which are discussed in some detail in \cite{Aluie11a}.
These represent the only two processes capable of direct transfer of kinetic energy \emph{across} scales.
Pressure dilatation, $-\OL{P}_\ell\grad\bdot\OL\bu_\ell$, does not contain any modes at scales $<\ell$ 
(or a moderate multiple thereof), at least for a filter kernel $\hat{G}(\bk)$ compactly supported in Fourier space.
Therefore, pressure dilatation cannot participate in the inter-scale transfer of kinetic energy and only 
contributes to conversion of large-scale kinetic energy into internal energy. 
This was a crucial observation made in \cite{Aluie11c} on which the analysis in this Letter will
be based.

\section{Pressure dilatation cospectrum} \label{sec:PressDil}
Kinetic and internal energy budgets couple through two mechanisms.
The first is viscous dissipation which was proved in \cite{Aluie11a} to be 
confined to the smallest scales $\ell\le \ell_\mu$ (the dissipation scale-range). Therefore, 
large-scale kinetic energy in (\ref{largeKE}) does not couple to internal energy 
via viscous dynamics for $\ell\gg\ell_\mu$. The second mechanism is pressure dilatation,
$-P\grad\bdot\bu$, which exchanges kinetic and internal energy via
compression and rarefaction.

It was shown in \cite{Aluie11c} that if pressure dilatation co-spectrum, defined  as
\be E^{PD}(k)\equiv \sum_{k-0.5<|\bk|<k+0.5} -\hat{P}(\bk)\widehat{\grad\bdot\bu}(-\bk),
\lb{PDcospectrumdef}\ee
decays fast enough as a function of wavenumber, 
\be |E^{PD}(k)| \le C\, u_{rms}\,P_{rms} \,(kL)^{-\beta},   \,\,\,\,\,\,\,\,\,\,\,\,\,\,\,\, \beta > 1,
\lb{scaling4}\ee
then mean pressure dilatation exchanges mean kinetic and internal energy
over a transitional ``conversion'' scale-range of limited extent. Here, $C$ is a dimensionless constant 
and $L$ is an integral scale. At smaller scales beyond the conversion
range, mean kinetic and internal energy budgets statistically decouple giving rise to 
an inertial range over which mean kinetic energy undergoes a scale-local conservative 
cascade. In this Letter, we provide the first empirical evidence in support of assumption 
(\ref{scaling4}) in Figure \ref{Fig:isoeosSpecPD} which we shall 
discuss more below.

The idea behind assumption (\ref{scaling4}) is straightforward and rests on the convergence
of a series or an integral at infinity. In the limit of large Reynolds number, assumption 
(\ref{scaling4}) implies that mean large-scale pressure dilatation, 
$PD(\ell)\equiv-\langle\OL{P}_\ell\grad\bdot\OL\bu_\ell\rangle$, asymptotes to a finite 
constant, $\theta\equiv-\langle P \grad\bdot\bu\rangle$, as $\ell\to 0$.
In other words, $PD(\ell)$
acting at scales $>\ell$ converges and becomes independent of $\ell$ at small enough scales:
\be
\lim_{\ell\to 0}PD(\ell) = \lim_{K \to \infty} \sum_{0<k<K} E^{PD}(k) = \theta.
\lb{PDresult}\ee
Note that $PD(\ell)$ in (\ref{PDresult}) (also shown in Figure \ref{Fig:isoeosSGSFlux}) 
is a cumulative quantity representing the contribution from
all wavenumbers $k< K=\ell^{-1}$, whereas spectra $E^{PD}(k)$ and $E^{u}(k)$ in 
Figures \ref{Fig:isoeosSpectra}, \ref{Fig:isoeosSpecPD} are density functions.
Convergence of  $PD(\ell)$ in (\ref{PDresult}) 
expresses the decoupling (in an average sense) between large-scale kinetic
and internal balances. Such a decoupling is statistical and does 
not imply that small scales evolve according to incompressible dynamics. 
However, while small-scale compression and rarefaction can still take place 
pointwise, they yield a vanishing contribution to the \emph{space-average}. 

\begin{figure}
\centering
\includegraphics[angle=0,totalheight=.4\textheight,width=.45\textwidth]{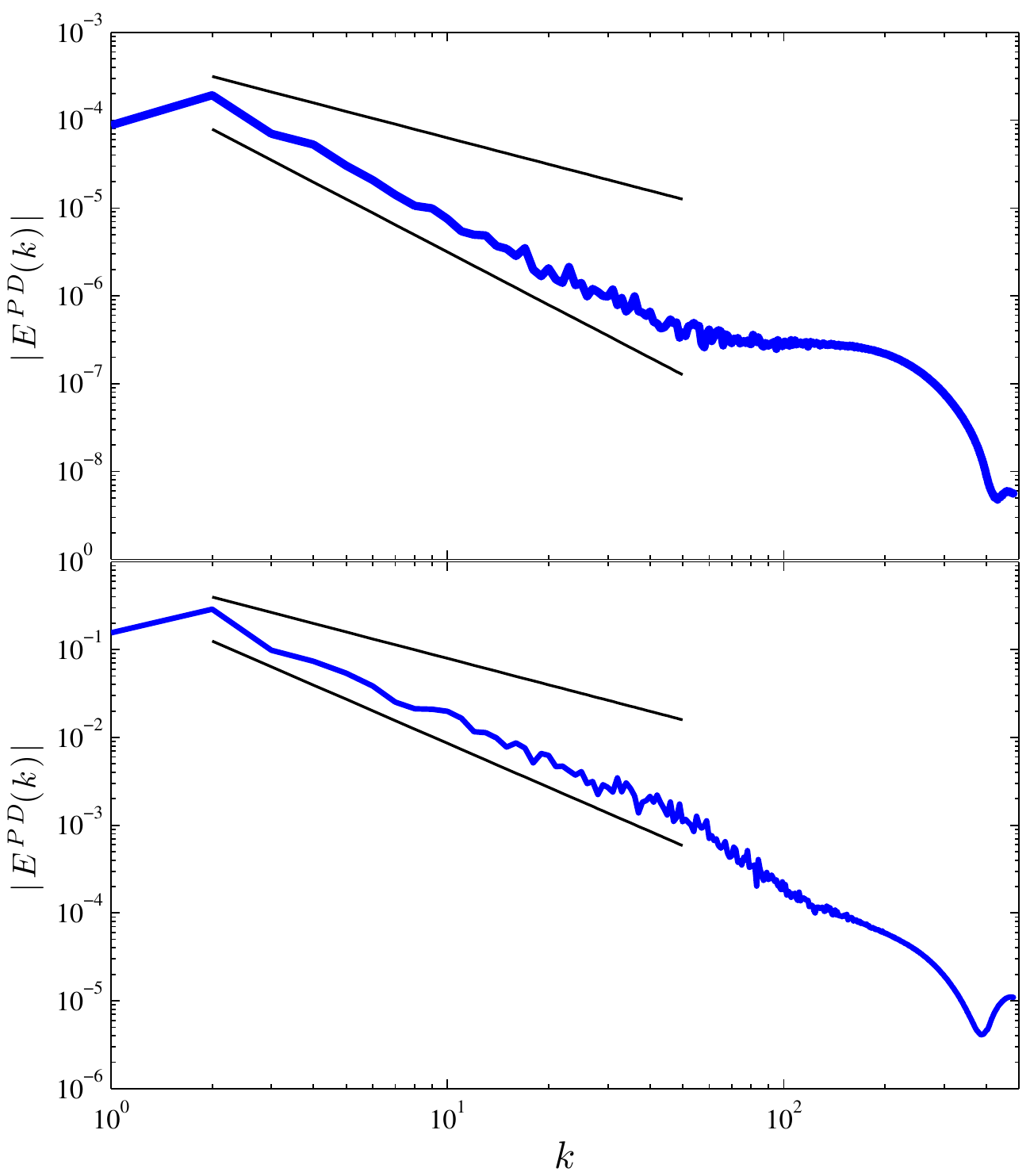}
\caption{Magnitude of pressure dilatation cospectrum from run III (top) and run IV (bottom). 
Straight lines are for reference and extend over the fitting range.
Top: straight lines have slopes of $-1$ and $-2$; power-law fit of data gives a slope of -1.88.
Bottom: straight lines have slopes of $-1$ and $-5/3$; power-law fit of data gives a slope of -1.61.}
\lb{Fig:isoeosSpecPD}\end{figure}

\begin{figure}
\centering
\includegraphics[angle=0,totalheight=.4\textheight,width=.45\textwidth]{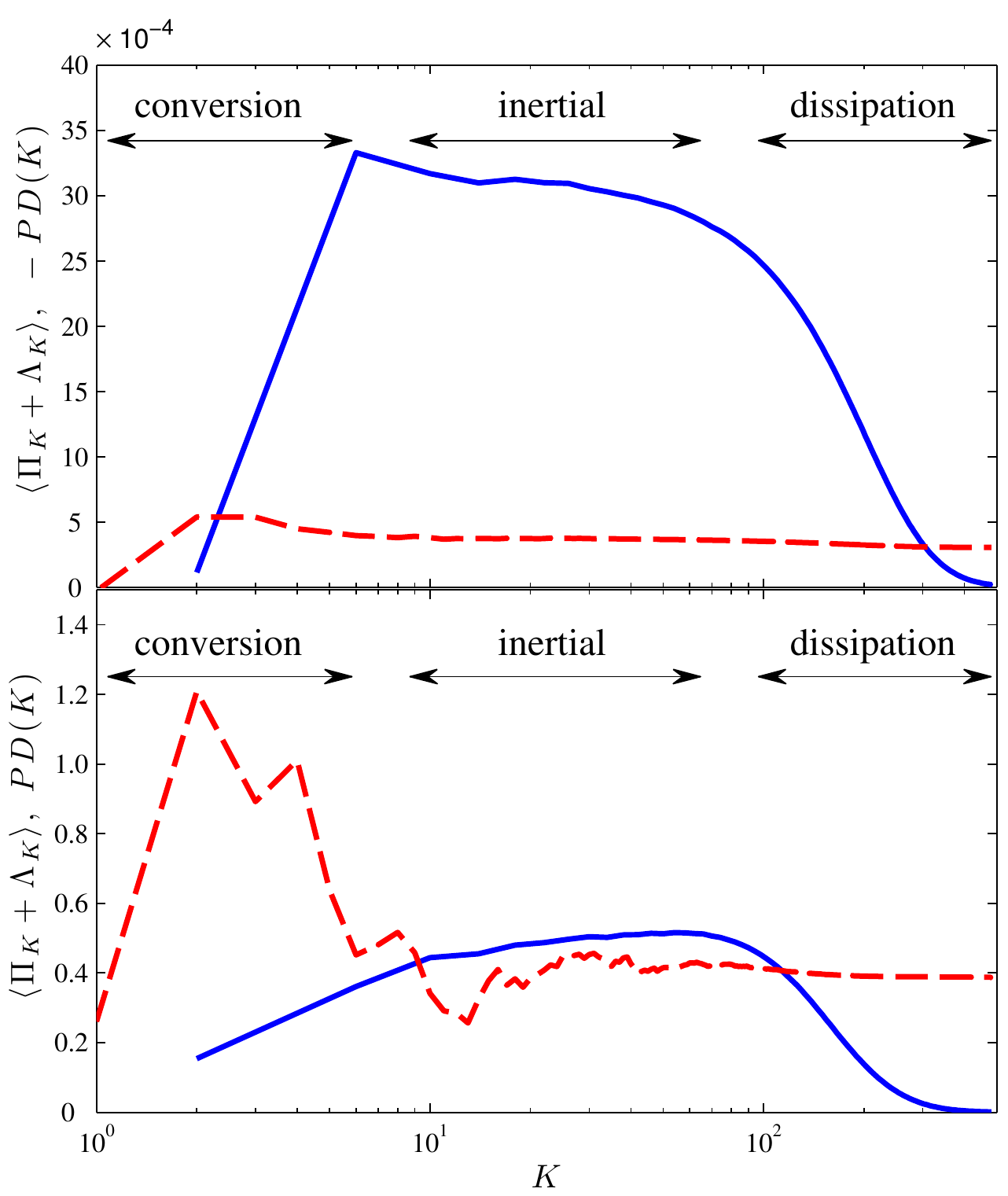}
\caption{
Solid (blue) plot shows mean SGS kinetic energy flux, $\langle \Pi_K + \Lambda_K\rangle$, 
and dashed (red) plot shows mean pressure dilatation, $PD(K)$, from run III (top) and run IV (bottom).
}
\lb{Fig:isoeosSGSFlux}\end{figure}

We denote the smallest wavenumber at which such statistical decoupling occurs by $K_c$. 

Over the ensuing wavenumber-range, $K_c<K\ll K_\mu$ (where $K_\mu=\ell_\mu^{-1}$), 
net pressure dilatation does not play a role. For $\ell=K^{-1}$ over this wavenumber-range,
budget (\ref{largeKE}) of large-scale kinetic energy becomes 
\be \langle\Pi_\ell+\Lambda_\ell \rangle  = \langle\epsilon^{inj}\rangle - \theta
\lb{inertialKEbudget}\ee
in a steady state and after space-averaging. 
If $\langle\epsilon^{inj}\rangle$ is localized to the largest scales as discussed in \cite{Aluie11a}, 
then $\langle \Pi_\ell+\Lambda_\ell\rangle$ will be a constant, independent of scale $\ell$.

A constant SGS flux implies that mean kinetic energy cascades conservatively to smaller scales,
despite not being an invariant of the governing dynamics. This was one of the main conclusions
in \cite{Aluie11c}. In particular, kinetic energy  can only reach dissipation scales via the SGS flux, 
$\Pi_\ell+\Lambda_\ell$, through a scale-local cascade process. We are therefore justified in calling wavenumber-range $K_c < K \ll K_\mu$ the inertial range of compressible turbulence. The existence
of this inertial range in such flows warrants expectations that spectra with power-law scalings 
should exist and that their observation is evidence of a turbulent cascade process similar 
(although not necessarily identical) to that in incompressible flows.

\section{Results} \label{sec:Results}
We will now present results from our simulations to provide the first empirical 
evidence in support of assumption (\ref{scaling4}). 
Figure \ref{Fig:isoeosSpecPD} shows that $|E^{PD}(k)|$ 
decays at a rate significantly faster than $k^{-1}$, well in agreement with condition (\ref{scaling4}).
We fitted the measurements with power laws over wavenumber range $k\in[2,50]$. 
We obtain a scaling of $|E^{PD}(k)|\sim k^{-1.88}$  from run III and $|E^{PD}(k)|\sim k^{-1.61}$ from run IV.
The measurements shown from runs III and IV are averaged over 21 and 28 evenly-spaced time 
snapshots, respectively. The plots of $|E^{PD}(k)|$ from run IV are normalized  at each time by the value 
of $\sum_k|E^{PD}(k)|$ at that time. We have verified that results from our  $512^3$ runs I and II are
 consistent with those presented.

To further illustrate the main idea of this Letter, we analyzed the SGS flux terms,
$\langle \Pi_\ell + \Lambda_\ell\rangle$, in budget (\ref{largeKE}).
Figure \ref{Fig:isoeosSGSFlux} shows that pressure dilatation, $PD(\ell)$,
tends to a constant beyond a transitional conversion range. Since $PD(\ell)$ is a cumulative 
quantity representing the contribution from all wavenumbers $k< K=\ell^{-1}$, its convergence 
to a constant beyond $K=10$ implies that any smaller scales with $k>10$ give a negligible 
contribution to pressure dilatation. The SGS kinetic energy flux
also becomes approximately constant\footnote{Unlike the cumulative quantity $PD(\ell)$
which represents the contribution from all wavenumbers $k< K=\ell^{-1}$, the SGS flux
$\langle \Pi_\ell + \Lambda_\ell\rangle$ is a hybrid quantity involving both wavenumbers
$k< K=\ell^{-1}$ (large-scale terms) and $k> K=\ell^{-1}$ (subgrid terms) 
as definitions (\ref{flux1}),(\ref{flux2}) show.}
 beyond this conversion range, implying that kinetic energy
is cascading conservatively (without any ``leakage'' to/from internal energy) until the dissipation range is reached. 
 It is notoriously hard
to achieve a constant SGS flux range in limited resolution simulations due to viscous contamination.
Even the largest-to-date simulation of incompressible turbulence on a $4096^3$ grid by \cite{Kanedaetal03} 
does not exhibit a clear constant flux. However, the plots in Figure \ref{Fig:isoeosSGSFlux} suggest that
an inertial range over which SGS flux is constant does indeed arise at smaller scales over which $PD(\ell)$ plateaus.

\section{The role of shocks} \label{sec:Shocks}
Our results and the physical picture we are advancing might seem counter-intuitive at first.
After all, a hallmark of compressible turbulence is the formation of shocks and the generation
of sound waves. Such phenomena involve compression and rarefaction at all scales and are not
restricted to small wavenumbers.

However, there is no contradiction between the existence of such phenomena and our conclusions.
Our results concern global pressure dilatation, $-\langle P\grad\bdot\bu\rangle$,
and not the pointwise quantity. While we expect very large pressure dilatation values in the 
vicinity of small-scale shocks, our results indicate that such a contribution will vanish
when averaging over the flow domain due to cancellations between compression and rarefaction
regions. For example, linear (small-amplitude) sound waves yield zero mean pressure dilatation, 
$-\langle P\grad\bdot\bu\rangle =0$. Figure \ref{Fig:Visualize} shows that values  of
pointwise pressure dilatation at small scales, $-[P\grad\bdot\bu(\bx)-\OL{P}_{\ell}\grad\bdot\OL\bu_{\ell}(\bx)]$,
are an order of magnitude more intense relative to large-scale pressure dilatation,
$-\OL{P}_{\ell}\grad\bdot\OL\bu_{\ell}(\bx)$, as expected. The distribution of 
$-[P\grad\bdot\bu(\bx)-\OL{P}_{\ell}\grad\bdot\OL\bu_{\ell}(\bx)]$ has heavy tails implying 
 spatially rare but intense two-way exchange between kinetic and internal energy. 
Furthermore, the small-scale distribution has positive skewness implying that conversion from kinetic into 
internal energy through compression occupies less volume and is more intense relative to conversion from internal into kinetic energy through rarefaction. Despite the larger intensity of conversion at small scales, 
its global contribution,  
$-\langle P\grad\bdot\bu-\OL{P}_{\ell}\grad\bdot\OL\bu_{\ell}\rangle = 2.0\times 10^{-6}$,
is negligible compared to the mean of large-scale conversion,
$PD(\ell)=-\langle \OL{P}_{\ell}\grad\bdot\OL\bu_{\ell}\rangle = 3.7\times 10^{-5}$,
for $\ell^{-1}=K=20$. This is consistent with the plateau of $PD(\ell)$ in Figure \ref{Fig:isoeosSGSFlux}.

The origin of such cancellations at small-scales 
may be understood through the following argument (\cite{Aluie11c}). 
While pressure in $-\langle P\grad\bdot\bu\rangle$ 
derives most of its contribution from the largest scales, $\grad\bdot\bu$ is dominated by the 
smallest scales in the flow. Therefore, pressure varies slowly in space, primarily at scales $\sim L$,
while $\grad\bdot\bu$ varies much more rapidly, primarily at scales $\ell_\mu\ll L$, leading to a 
decorrelation between the two factors.
Our numerical results presented in this Letter can be regarded 
as empirical support to such a physical argument.

\begin{figure}
\centering
\includegraphics[angle=0,totalheight=.7\textheight,width=1\textwidth]{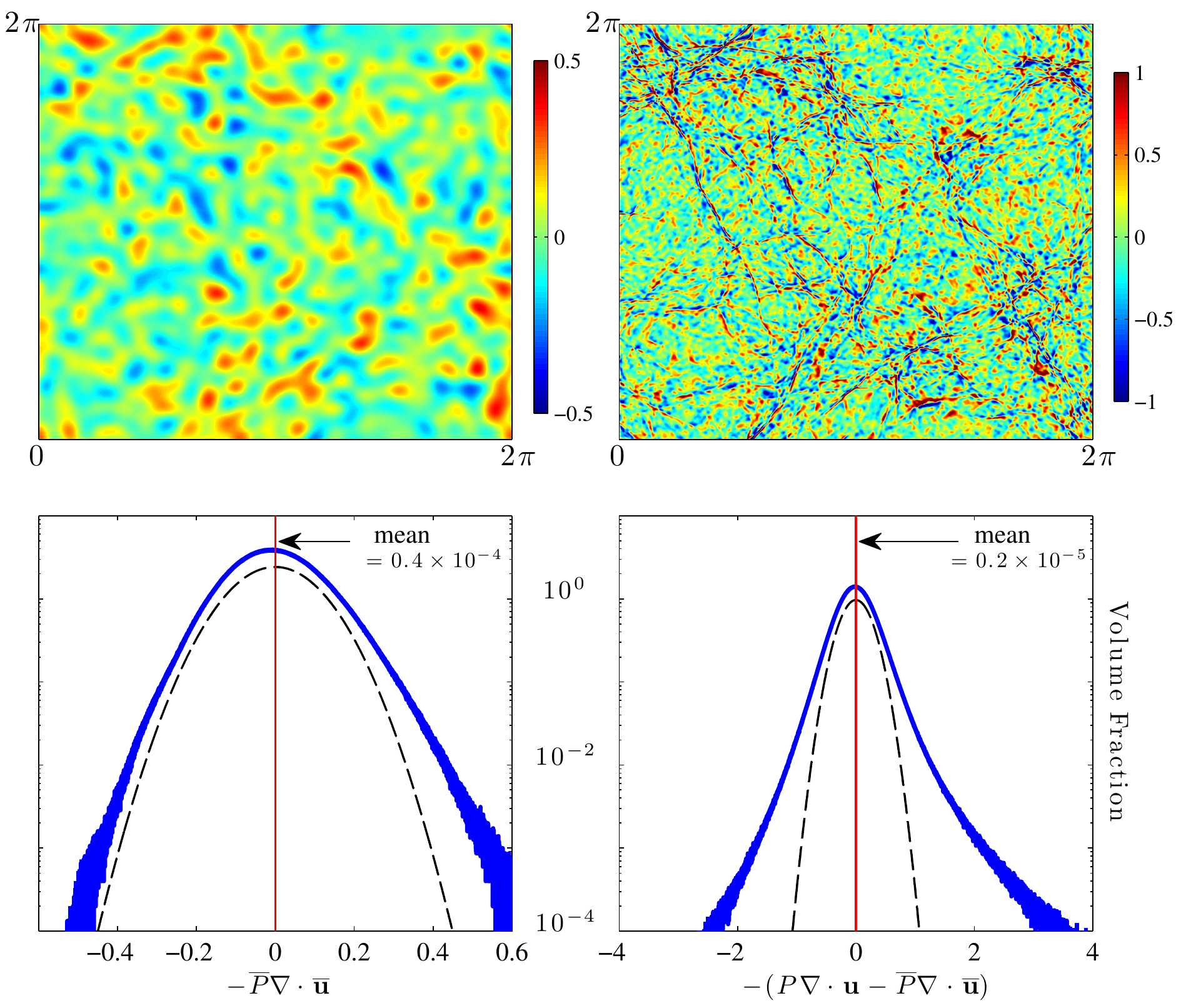}
\caption{For $\ell^{-1}=K=20$ in Fig. \ref{Fig:isoeosSGSFlux} from run IV, 
top two panels visualize pointwise pressure dilatation at large scales (top-left), 
$-\OL{P}_{\ell}\grad\bdot\OL\bu_{\ell}(\bx)$,
and the residual from small-scales (top-right), 
$-P\grad\bdot\bu(\bx)+\OL{P}_{\ell}\grad\bdot\OL\bu_{\ell}(\bx)$, in a 
$1024^2$ domain cross-section. The color maps are rescaled to fit within the color bars' bounds. 
Bottom two panels show the distributions of large-scale (bottom-left) and small-scale (bottom-right)
pressure dilatation. Dashed curves are for reference and show unnormalized gaussian distributions 
with zero mean.
}
\lb{Fig:Visualize}\end{figure}

 \section{Conclusions \& Discussion} \label{sec:Conclusion}
The main result of our Letter was to provide the first empirical evidence
that pressure dilatation co-spectrum decays at a rate faster than $k^{-1}$,
in accord with condition (\ref{scaling4}). This is sufficient to imply that exchange
of kinetic and internal energy through compression and rarefaction takes place
at the large scales, on average. At smaller scales beyond a transitional 
``conversion'' scale-range, mean kinetic and internal energy budgets statistically 
decouple and kinetic energy can only reach dissipation scales via a scale-local
conservative cascade process. The existence of such a conservative cascade in 
compressible turbulence justifies expectations that spectra with power-law scalings 
should exist in such flows. Furthermore, it suggests that  observations of power-law
spectra in astrophysical systems are indeed evidence of a turbulent cascade process 
similar (although not necessarily identical) to that in incompressible flows.

This work leads us to some new questions which we hope to address in future studies.
Does the decay rate of pressure dilation cospectrum depend on the 
decaying/forced nature of the turbulence or on other parameters of the flow
such as equation of state, Mach number, and the ratio of compressive-to-solenoidal 
components of the velocity field? Is the power law scaling of pressure dilation cospectrum
that we find in Figure \ref{Fig:isoeosSpecPD} reflecting the asymptotic scaling
at arbitrarily high Reynolds numbers? Does the value of $PD(\ell)$ in
Figure \ref{Fig:isoeosSGSFlux} relative to that of SGS flux increase by increasing the
compressive amplitude of the forcing?
We invite future studies to investigate these questions and to try and reproduce our
results presented in this Letter. Verifying assumption (\ref{scaling4}) 
under a variety of controlled conditions would substantiate the idea of statistical
decoupling between kinetic and internal energy budgets. This is of paramount
importance for future attempts to extend Kolmogorov's ideas on a conservative
cascade of kinetic energy to compressible turbulence.

\vspace{0.4cm}
\noindent {\small
{\bf Acknowledgements.} 
We are grateful to J. Cho for providing us with his forcing subroutine.
H. A. thanks R. E. Ecke, G. L. Eyink, S. S. Girimaji, S. Kurien, and D. Livescu
for useful discussions. H.A. acknowledges partial support from NSF grant
PHY-0903872 during a visit to the Kavli Institute for Theoretical Physics.
This research was performed under the auspices of the 
U.S. DOE at LANL under Contract No. DE-AC52-06NA25396
and supported by the LANL/LDRD program
and by the DOE/Office of Fusion Energy Science through
NSF Center for Magnetic Self-Organization.}

\bibliographystyle{unsrt}

\end{document}